\documentclass[aip, reprint]{revtex4-1}
\usepackage[dvips]{color}
\usepackage{ulem}
\usepackage{xcolor}

\usepackage{hyperref}
\usepackage{amsmath}
\hypersetup{colorlinks=true, linkcolor=blue, filecolor=blue, urlcolor=blue, citecolor=blue}
\usepackage{graphicx}

\begin{document}

\title{Calibration of BAS-TR image plate response to GeV gold ions}

\author{D.~Doria}\email[Author to whom correspondence should be addressed: ]{domenico.doria@eli-np.ro}
\affiliation{Centre for Plasma Physics, School of Mathematics and Physics, Queen's University Belfast, BT7 1NN, UK}
\affiliation{Extreme Light Infrastructure (ELI-NP), and Horia Hulubei National Institute for R\&D in Physics and Nuclear Engineering (IFIN-HH), Str. Reatorului No. 30, 077125 Bucharest-Magurele, Romania}

\author{P.~Martin}
\affiliation{Centre for Plasma Physics, School of Mathematics and Physics, Queen's University Belfast, BT7 1NN, UK}

\author{H.~Ahmed}
\affiliation{Centre for Plasma Physics, School of Mathematics and Physics, Queen's University Belfast, BT7 1NN, UK}
\affiliation{Central Laser Facility, Rutherford Appleton Laboratory, Didcot, Oxfordshire, OX11 0QX, UK}
\author{A.~Alejo}
\affiliation{Centre for Plasma Physics, School of Mathematics and Physics, Queen's University Belfast, BT7 1NN, UK}
\affiliation{IGFAE, Universidade de Santiago de Compostela, Santiago de Compostela, Spain}

\author{M.~Cerchez}
\affiliation{Institut f\"ur Laser-und Plasmaphysik, Heinrich-Heine-Universit\"at, D\"usseldorf, Germany}

\author{S.~Ferguson}
\affiliation{Centre for Plasma Physics, School of Mathematics and Physics, Queen's University Belfast, BT7 1NN, UK}

\author{J.~ Fernandez-Tobias}
\affiliation{Central Laser Facility, Rutherford Appleton Laboratory, Didcot, Oxfordshire, OX11 0QX, UK}
\affiliation{Instituto de Fusion Nuclear, Universidad Polit\'ecnica de Madrid, Madrid, Spain}

\author{J.~S.~Green}
\affiliation{Central Laser Facility, Rutherford Appleton Laboratory, Didcot, Oxfordshire, OX11 0QX, UK}

\author{D.~Gwynne}
\affiliation{Centre for Plasma Physics, School of Mathematics and Physics, Queen's University Belfast, BT7 1NN, UK}

\author{F.~Hanton}
\affiliation{Centre for Plasma Physics, School of Mathematics and Physics, Queen's University Belfast, BT7 1NN, UK}

\author{J.~Jarrett}
\affiliation{Department of Physics, SUPA, University of Strathclyde, Glasgow G4 0NG}

\author{D.~A.~Maclellan}
\affiliation{Department of Physics, SUPA, University of Strathclyde, Glasgow G4 0NG}

\author{A.~McIlvenny}
\affiliation{Centre for Plasma Physics, School of Mathematics and Physics, Queen's University Belfast, BT7 1NN, UK}

\author{P.~McKenna}
\affiliation{Department of Physics, SUPA, University of Strathclyde, Glasgow G4 0NG}


\author{J.A.~Ruiz}
\affiliation{Instituto de Fusion Nuclear, Universidad Polit\'ecnica de Madrid, Madrid, Spain}

\author{M.~Swantusch}
\affiliation{Institut f\"ur Laser-und Plasmaphysik, Heinrich-Heine-Universit\"at, D\"usseldorf, Germany}

\author{O.~Willi}
\affiliation{Institut f\"ur Laser-und Plasmaphysik, Heinrich-Heine-Universit\"at, D\"usseldorf, Germany}

\author{S.~Zhai}
\affiliation{ELI Beamlines, Za Radnicí 835, Dolní Břežany, 252 41, Czech Republic}

\author{M.~Borghesi}
\affiliation{Centre for Plasma Physics, School of Mathematics and Physics, Queen's University Belfast, BT7 1NN, UK}

\author{S.~Kar}\email{s.kar@qub.ac.uk}
\affiliation{Centre for Plasma Physics, School of Mathematics and Physics, Queen's University Belfast, BT7 1NN, UK}

\date{\today}

\begin{abstract}

The response of the BAS-TR image plate (IP) was absolutely calibrated using CR-39 track detector for high linear energy transfer (LET) Au ions up to $\sim$1.6 GeV (8.2 MeV/nucleon), accelerated by high-power lasers. The calibration was carried out by employing a high-resolution Thomson parabola spectrometer, which allowed resolving Au ions with closely spaced ionization states up to 58$^+$. A response function was obtained by fitting the photo-stimulated luminescence (PSL) per Au ion for different ion energies, which is broadly in agreement with that expected from ion stopping in the active layer of the IP. This calibration would allow quantifying the ion energy spectra for high energy Au ions, which is important for further investigation of the laser-based acceleration of heavy ion beams.

\end{abstract}

\pacs {}

\maketitle

\section{Introduction}

Laser-driven ion acceleration has attracted considerable interest over the last two decades. Thanks to the ongoing development in high-power laser technology, advanced interaction mechanisms, such as radiation pressure acceleration (RPA)~\cite{Qiao2012,Kar2012}, and relativistically induced transparency (RIT) enhanced acceleration~\cite{Higginson2018}, allow for an efficient acceleration of heavy ion species from the bulk of the laser-irradiated thin foil targets. Laser-acceleration of high-density bunches of heavy-ions (such as gold ions) to tens of MeV/nucleon is of particular interest for applications in nuclear physics and high-energy density physics (HEDP)~\cite{Habs2011}, and has gained popularity in recent years because of this~\cite{Domanski2018,Lindner2019, Wang2021, Petrov2016, Petrov2017}.

Since the ion beams produced by high power lasers are usually multi-species~\cite{Macchi2013,Kar2012}, Thomson parabola spectrometers (TPS)~\cite{Gwynne2014} are deployed to characterize them. TPS assembly is coupled to a range of detectors, passive or active, such as, solid state detectors (e.g. CR-39), scintillators, micro-channel plate (MCP) and image plates (IPs). The IPs are widely used in laser-plasma experiments due to their non-disposable nature (i.e., the signal on the IP can be erased after every use), high dynamic range ($\sim$10$^5$), and high spatial resolution (up to 10 $\mu$m). In order to make the IP an absolute ion flux detector, different types of IPs have been absolutely calibrated for several ion species (e.g., protons~\cite{Mancic2008}, deuterium~\cite{Alejo2014}, carbon~\cite{Doria2015}, titanium~\cite{Strehlow2019}, and silver and xenon ions~\cite{Nishiuchi2020}) by using laser-driven ion beams. In light of the emerging acceleration mechanisms that can accelerate heavy ions efficiently, it is crucial to calibrate the response of IPs to energetic heavier ions.

In this article, we report the response of the BAS-TR image plate to gold ions of energies up to $\sim$1.6 GeV. The calibration was obtained deploying a slotted CR-39 nuclear track detector in conjunction with the IP, in a technique similar to the one used in ref’s.~\cite{Alejo2014, Doria2015}. This method enabled a direct comparison between the signal on IP and the number of particles (in this case, Au ions) on CR-39, for a wide range of energies and charge states.

\begin{figure*}[t]
	\includegraphics[width=\textwidth]{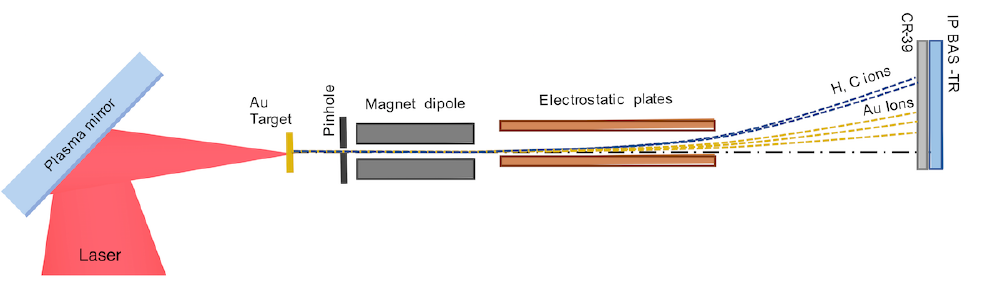}
	\caption{A schematic of the experimental setup employed at the Vulcan Petawatt (VPW) target area. The drawing shows a detailed sketch of TPS deployed in the experiment, depicting the regions of the static magnetic and electric fields for energy dispersion and separation of different ion species. A slotted grid of CR-39 was placed in front of the IP to cross-calibrate its response to Au ions.}
	\label{setup_gold}
\end{figure*}

\section{Experimental setup}

The data shown in this manuscript was obtained from two experimental campaigns, performed under effectively same experimental conditions, using the petawatt arm of the Vulcan laser system at the Rutherford Appleton Laboratory (RAL), STFC, UK. The laser beam in both experiments was with a pulse duration of $\sim$850 fs and it was  focused onto ultra-thin (tens of nanometre) Au foils by an \textit{f}/3 off-axis parabolic mirror to a spot size of $\sim$5 $\mu$m full width half maximum (FWHM). A plasma mirror was deployed in both experiments before the target, in order to improve the temporal contrast of the incident laser pulse, by suppressing unwanted pre-pulses and the amplified spontaneous emission that arises during the amplification process, known as the pedestal. After reflection from the plasma mirror, the energy on the target in both experiments was $\sim$200 J, with 35$\%$ of the energy in the focal spot, which delivered an intensity of $\sim$3--5$\times$10$^{20}$ W/cm$^{2}$ on target. As shown in Fig.~\ref{setup_gold}, a Thomson Parabola Spectrometer (TPS) assembly employing image plate (IP) and CR-39 as detectors, similar to the setup described in~\cite{Gwynne2014, Alejo2014}, was used in both experiments to characterize the energy distribution of multi-species ion beams generated in the intense laser interaction with ultra-thin foils.

CR-39 is a plastic polymer used as a solid-state track detector in nuclear science to detect ions and neutrons~\cite{Kar2007a, Kanasaki2016, Phillips2006Sep}. In this experiment, 1 mm thick sheets of CR-39 with rectangular slots of 1.5 mm high and bars of 2 mm high were used. The CR-39 was placed on top of the IP, which were then both wrapped in 6 $\mu$m thick aluminum foil. This is to protect the IP from unavoidable and undesired radiations emitted from the intense laser-plasma interaction, which may worsen the signal-to-noise level and reduce the signal by stimulated emission. The foil also reduces the energy of the gold ions impinging on the detector, and the energy loss was accounted for using the Monte Carlo SRIM simulations~\cite{SRIM, Ziegler2010Jun} to determine the energy of Au ions incident on the IP active layer. After the shot, the exposed IPs (BAS-TR) were scanned by an image plate scanner (Fujifilm FLA-5000)~\cite{IP}.

\section{Imaging Plate Detector}

There are different types of IPs, and they are widely used as a detector of ionizing radiation such as X-rays, electrons, and ions. The particle is registered in its active layer made of grains of barium fluoride phosphor doped with Europium which get stimulated and stores the energy deposited by the impinging particles. The signal is retained in a metastable state for several hours while slowly decaying and can be released on demand via the photo-stimulated luminescence (PSL) phenomenon. For the data described in this paper, we have used the Fujifilm BAS-TR, which does not have a protective coating in front of the active layer, and thus enables the detection of low-energy ions.

As mentioned earlier, the Fujifilm FLA-5000 image plate scanner was used to read the IP. Each pixel of the image stores the energy emitted via PSL in a 16 bit digitized Quantum levels ($Q_L$), which can be converted to PSL value by using the following formula~\cite{IP},

\begin{equation}\label{eq:PSLscanner}
\mbox{PSL$ = \left ( \frac{R}{100} \right )^2 \times \frac{4000}{S} \times 10^{L\left(\frac{QL}{2^G -1} - \frac{1}{2}\right)}$}
\end{equation}

\noindent where $R$ is the scanner resolution that is the pixel size of the RAW image in microns, $S$ is the scanner sensitivity that is related to the voltage applied to the PM, $L$ is called latitude and corresponds to the dynamic range chosen for the signal digitization, and $G$, normally set to 16, is the bit depth of the image. 

The signal level on the IP fades out over time due to spontaneous decay. Therefore, the signal must be corrected for the time elapsed between irradiation and scanning. To make any calibration a standard, which can be used for any experimental data, it is therefore necessary to set a certain scanning time as a reference one to which IP scan can be normalized.
As the fading slows down significantly after 20--30 minutes after exposure~\cite{Alejo2014}, the PSL signal was corrected to a 30 min reference time after the exposure by using the empirical formula described in refs.~\cite{Alejo2014,Doria2015}. 

\begin{equation}\label{eq:PSL30timedecay}
\mathrm{PSL_{30} }= \left ( \frac{30}{t} \right )^{-0.161} \times \mathrm{PSL(t)}
\end{equation}

\noindent where PSL$_{30}$ is the PSL value re-normalized to 30 minutes after exposure, and $t$ is the time in minutes between exposure and scanning.

\subsection{Calibration of IP response to Au ions}

In order to calibrate the response of the IP for Au ions, a slotted CR-39 was used in conjunction with IPs to relate the PSL at a given energy measured by the IP to the number of incident ions, measured absolutely by the CR-39. The CR-39 was etched in a 6M sodium hydroxide solution kept at 85 $^{\circ}$C for several minutes. The CR-39 was checked under a microscope periodically so that over-etching (pit size increasing to the point of overlapping each other) did not occur.

\begin{figure}[h]
	\centering
	\includegraphics[width=0.45\textwidth]{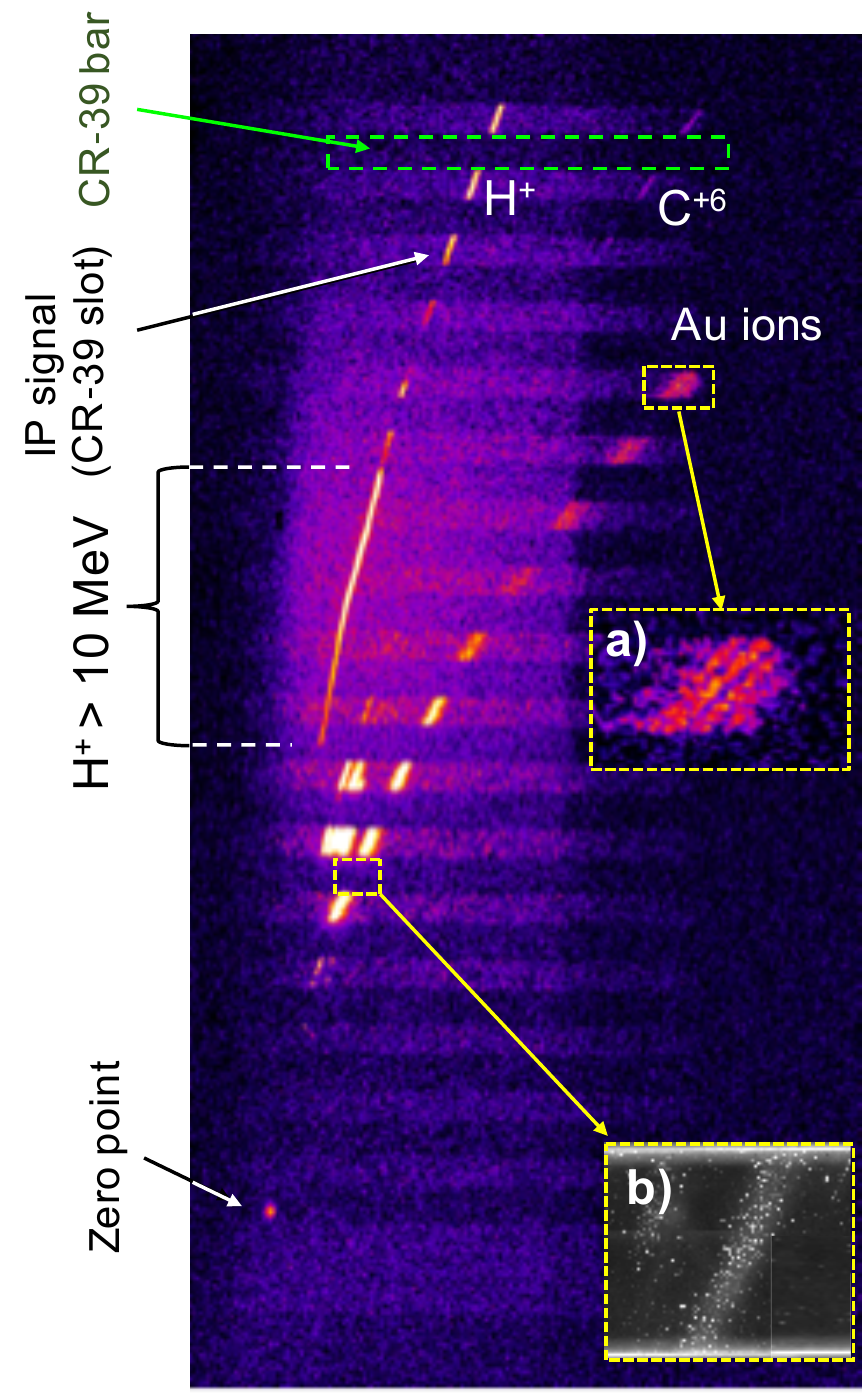}
	\caption{Raw image of IP signal for a shot: the shadow of the slotted CR-39 is clearly visible on the IP. Ion energy increases going from top to bottom on the image, towards the zero point. The inset (a) shows well separated traces of Au ions of neighbouring charge states around 58+, while the inset (b) shows an example of pits appeared after etching the CR-39 in a location where gold ions impinged. The background around the signal is mainly due to the X-ray produced by the laser interaction and the scattered ions passing through the pinhole edge, while the low backgrounds on left and right are caused by clipping of the extraneous signals by the magnets and electric plates of the TPS. }
	\label{IP_CR39}
\end{figure}

Fig.~\ref{IP_CR39} shows the scanned IP image after PSL conversion. Clear ion traces are visible, corresponding to proton, carbon, and gold ions. The shadow of the slotted CR-39 placed over the IP is also clear, which blocks the traces of the heavier ions and lower energy ($<$10 MeV) protons. Inset (a) shows a zoomed in section of the IP, showing a low energy region of the Au ion traces. There are several noticeable smaller traces comprising the entire Au signal, indicating that a range of charge states, up to a maximum of 58+, were formed during the laser pulses interaction. Inset (b) shows a section of the slotted CR-39 imaged under a microscope at $\times$10 magnification, towards the high energy region of the gold ion trace. The pits formed by Au ions can be clearly seen as the streak of white spots illuminated by the microscope.

After etching, the gold pits at each location on the CR-39 were imaged using a microscope and manually counted. They were then sorted into bins of equal area as the pixel size of the IP image (25 $\mu$m $\times$ 25 $\mu$m). 
The PSL per particle was then evaluated by summing the PSL bins close to a CR-39 edge, across the whole Au trace width (i.e. integrating across all charge states), and doing the same for the CR-39 pits in the bins at the edge of each slot (Fig.~\ref{IP_CR39}(a) and (b)). The relationship between the PSL and the ion pits at the edges was found by assuming valid a local linear extrapolation of the spectrum between a CR-39 edge and the IP. The uncertainty on the energy arises mainly from the range of ion charge state inside the binning, as at any CR-39 edge, ions of different charge state have different energy because dissimilarly deflected by the TPS.
The uncertainty on the response is mainly coming from the uncertainty on the PSL signal, which is due to the noisiness of the background. Due to the non-uniformity in the background level across the IP, background subtraction was carried out by using the background adjacent to the ion track. Whereas, the average background can be evaluated and subtracted, the existing noise generates a non-zero variance of the mean background value which is the primary source of uncertainty in the PSL. The error in the measurement of PSL signal goes from $\sim$7$\%$ for high energy ions to $\sim$10$\%$ for low energy ions. The total error in PSL/ion was calculated by propagating the error in PSL signal with the error in counting the pits on CR39. The latter was in the range 5-8$\%$ across the energy range (higher uncertainty was for higher energies, mainly due to higher flux).

\begin{figure}[t]
	\centering
	\includegraphics[width=\columnwidth]{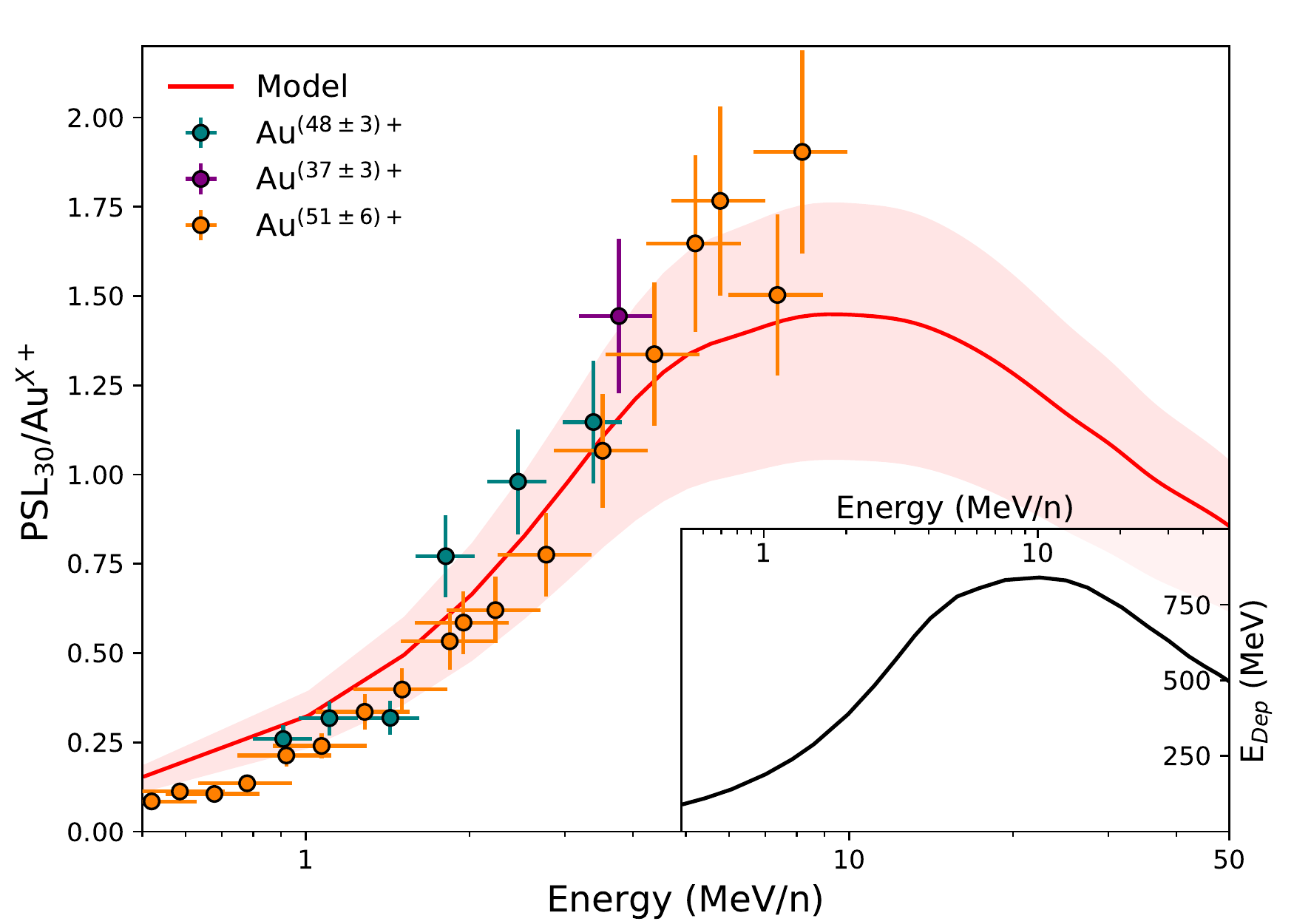}
	\caption{Absolute response of BAS-TR image plates to Au ions, showing the relation between the PSL (after 30 minutes) on the IP per incident Au ion for different Au energies and charge states. The data points show the experimental data from both campaigns (the orange points are from the first campaign, and the green and purple points are from the second one), errorbars represent a $\sim$10\% uncertainty in PSL and pits evaluation, and uncertainty in energy arising from the range of possible charge states. The solid curve is a non-linear least squares fitting of the data to the model described in Eq.~\eqref{Nishiuchi_model}. The red shaded areas represent upper (lower) bounds to the sensitivity parameter $\alpha_2$, by fitting only to the high (low) energy data points. The inset shows the total energy deposited per ion in the active layer by gold as a function of incident energy calculated using SRIM. The $x$ axis is the same in the inset as in the main figure.
	}
	\label{Calibration curve Au}
\end{figure}

Whilst calculating the energy for Au ions using Monte Carlo SRIM simulations, 6 $\mu$m aluminium foil covering the IP was considered. The  PSL$_{30}$/Ion for energies in the range of 0.5--8.2 MeV/n (corresponding to $\sim$0.1--1.6 GeV) was calculated, and the resulting data from both campaigns are plotted in Fig.~\ref{Calibration curve Au}. Considering the response curve, fluxes up to the order of 10$^7$ Ion/cm$^2$ of high energy (MeV/nucleon) Au ions can be detected with high accuracy using image plate detectors.

During each campaign, gold ions in a range of different charge states have been detected. However, it can be seen, as expected, that they all follow the same trend in their PSL response. This result is consistent with what has previously been observed~\cite{Doria2015, Strehlow2019, Nishiuchi2020}, as the ion stopping has been shown to not depend on the ion charge state, but only on initial kinetic energy, and its effective charge inside the material~\cite{Ziegler2010Jun, Betz1972Jul, Brandt1982May}.

Shown in the inset in Fig.~\ref{Calibration curve Au} is the energy deposited in the active layer, $E_{Dep}$, by gold ions of various energies. The total energy deposited could be calculated from the linear energy transfer (LET), $dE/dx$, given by SRIM simulations using active layer properties outlined by Bonnet \textit{et al.}~\cite{Bonnet2013Jan} by the integral:

\begin{equation}
\centering
E_{Dep} = \int_{0}^{w} \frac{dE}{dx} dx
\label{EnDep}
\end{equation}

\noindent where $w$ is the active layer thickness (50 $\mu$m for BAS-TR), and $x$ is the distance into the active layer the ion has traversed. The deposited energy is the primary factor that determines the IP response, and it can be seen that the energy deposition curve has a plateau region between $\sim$5--15 MeV/n, where $E_{Dep}$ does not change significantly. Thus, it may be assumed that the PSL response also does not vary across this range.

The response up to $\sim$8 MeV/n has been determined experimentally, however one can use models to fit the response, and therefore extrapolate to higher energies. The model described by Nishiuchi \textit{et al.}~\cite{Nishiuchi2020}, was adapted from previous models~\cite{Bonnet2013Jan, Lelasseux2020} to account for radiation with very high LET. They suggested a response function of the form:

\begin{equation}
\centering
PSL_0 = \int_{0}^{w} \frac{dE}{dx} \left ( \frac{\alpha_1 e^{-x/L}}{1 + kB\frac{dE}{dx} } + \alpha_2 \right ) dx
\label{Nishiuchi_model}
\end{equation}

\noindent where $dE/dx$, $x$, and $w$ are defined by Eq.~\eqref{EnDep}, $PSL_0$ is the PSL immediately after exposure, $L$ is a characteristic absorption length for the IP ($44\pm4$ $\mu$m for BAS-TR~\cite{Bonnet2013Oct}), $kB$ is a quenching factor (calculated for BAS-TR to be 0.15 Angstrom/eV~\cite{Lelasseux2020}), and $\alpha_1$ and $\alpha_2$ are sensitivity parameters, dependent on the type of incident radiation and IP.

For radiation with low stopping powers (e.g. electrons, protons) $\alpha_1 \gg \alpha_2$, and the first term dominates. However, for very large values of $dE/dx$, as is the case with GeV gold ions, $\alpha_2$ can become large compared to $\alpha_1$, and the second term dominates. As shown in Fig.~\ref{Calibration curve Au}, the model can be fitted reasonably well with the data points for $\alpha_2 = 2.81 \times 10^{-3}$ $ \mathrm{PSL_0/MeV}$.

The least squares fitting produces a value of $\alpha_1$ to be almost zero ($\sim$$10^{-13}$ $ \mathrm{PSL_0/MeV}$), although the error in this calculation (one standard deviation) was significantly larger, $\sim$1, and it was observed that using any value within this range made no discernible difference to the curve. The one standard deviation error in $\alpha_2$ was calculated as $\pm0.26 \times 10^{-3}$, however upper and lower bounds may also be calculated by considering a fitting appropriate to either lower and higher energy regions. These are shown by the shaded area in Fig.~\ref{Calibration curve Au}, and result in a final calculated range of the sensitivity for $\alpha_2 = (2.81^{+0.6}_{-0.8} ) \times 10^{-3}$ $ \mathrm{PSL_0/MeV}$.

Although the model broadly agrees with the experimental data points, there is scope for further development to fit the data points more closely across the entire energy range. At low energies, the range of the Au ions in the active layer is on the order of the grain size of the medium ($\sim$5--10 $\mu$m~\cite{Leblans2011Jun, Doria2015}) and a significant fraction of their energy may be lost to the resin that binds the phosphor. This would result in a lower response than what is predicted by the models. At high energies, close to the peak of the energy deposition curve and beyond, production of high energy (multi-MeV) secondary radiation (e.g. delta rays) could stimulate the phosphor in a column around the particle track, resulting in a higher measured response than what is predicted by the model. These two effects could lead to a ``steeper'' rise compared to what is predicted, which is what is observed experimentally. The interaction of extremely heavy, high LET particles (such as gold) with matter is a complex topic that requires further study, and is beyond the scope of this paper.

\begin{figure}[h]
	\centering
	\includegraphics[width=\columnwidth]{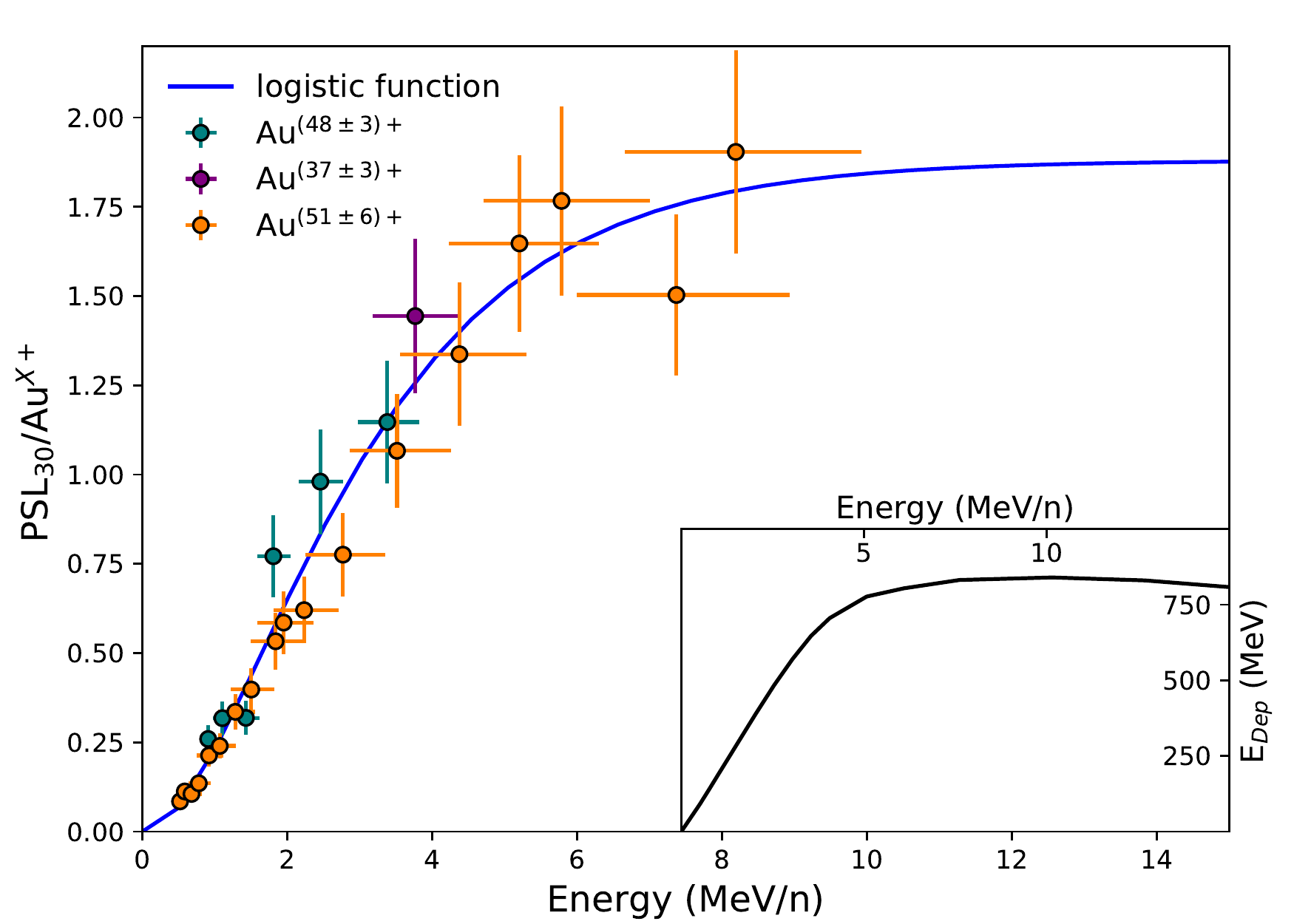}
	\caption{The same data as shown in Fig.~\ref{Calibration curve Au}, in a linear $x$ axis scale. The data has been fit with the logistic function described by Eq.~\eqref{logisticfit}. The inset again shows the total energy deposited in the active layer by gold ions as a function of incident energy, this time also in a linear $x$-axis scale, as in the main figure.
	}
	\label{fit_curve}
\end{figure}

To use the data points directly in data analysis, we can fit a function to the data, as has been done for previous IP calibrations~\cite{Mancic2008, Alejo2014, Doria2015}. As shown in the inset of Fig.~\ref{fit_curve}, the deposited energy up to 15 MeV/n resembles a sigmoid curve, with a plateau between about 7 and 14 MeV/n. Because in the plateau region the deposited energy is practically constant, the response would be expected to not change significantly. Therefore, the data has been fit using a sigmoid function and in particular a generalized logistic function. The fit is given by Eq.~\eqref{logisticfit}, where $E$ is the gold ion energy in units of MeV/nucleon. From the above-mentioned discussion, one can reasonably consider valid the extrapolation to energies up to $\sim$14 MeV/n (i.e., 2.75 GeV).

\begin{equation}
PSL_{30}/Au^{X+} = 1.88 \left ( 1 - \mathrm{e}^{-E/2.13} \right ) ^{2.14}
\label{logisticfit}
\end{equation}

\section{Conclusion}

The commonly used Fujifilm BAS-TR type image plate has been absolutely calibrated to Au ions for a range of energies ($\sim$0.5--8.2 MeV/n) by using Au ions accelerated by high-power lasers. The experimental data broadly agrees with a known model of IP response for high-LET particles. A logistic function has been fitted to the experimental data points, which, as per the energy deposition curve, can be reliably extrapolated up to 14 MeV/n.
This calibration will be useful in studies in laser-driven ion acceleration, in particular as interest in the acceleration of very heavy ions has peaked in recent years~\cite{Domanski2018, Lindner2019, Wang2021}, and the prospectives heavy ion acceleration in future multi-PW laser systems will see generation of Au beams of several GeV~\cite{Petrov2016, Petrov2017}.

\begin{acknowledgments}
We gratefully acknowledge funding from EPSRC, UK (No. EP/J002550/1-Career Acceleration Fellowship held by S.K., Nos. EP/K022415/1, EP/P010059/1, EP/R006202/1, and EP/J500094/1), and STFC, UK (No. ST/P000142/1). S.Z. acknowledges support by the Chinese Scholarship Council. The authors also acknowledge support from the members of the experimental science group, mechanical engineering, and target fabrication group of the CLF, STFC, UK. We would like to acknowledge the support, the helpful discussion and contribution to the paper by the late Prof. David Neely.

The data that support the findings of this study are available from the corresponding author upon reasonable request.

The authors have no conflicts to disclose.
\end{acknowledgments}

\bibliography{library}


\end{document}